\documentclass{JHEP3}
\title{AdS/CFT v.s. String Loops}

\author{Gianluca Grignani\\Dipartimento di Fisica and Sezione
I.N.F.N., Universit\`a di Perugia, Via A. Pascoli I-06123, Perugia,
Italia. \email{E-mail:grignani$@$pg.infn.it} }

\author{Marta Orselli \\NORDITA,  Blegdamsvej 17, DK-2100 Copenhagen, Denmark. \email{E-mail:orselli$@$nbi.dk} }

\author{Bojan Ramadanovic, Gordon W. Semenoff and Donovan Young\\Department of Physics and Astronomy,
University of British Columbia, 6224 Agricultural Road, Vancouver,
British Columbia V6T 1Z1 Canada.
\email{E-mail:bramadan,gordonws,dyoung$@$phas.ubc.ca}}

\abstract{ The one string-loop correction to the energies of two
impurity BMN states are computed using IIB light-cone string field
theory with an improved 3-string vertex that has been proposed by
 Dobashi and Yoneya. As in previous published computations, the
string vertices are truncated to the 2-impurity channel.  The result
is compared with the prediction from non-planar corrections in the
BMN limit of $\mathcal{N}=4$ supersymmetric Yang-Mills theory.  It
is found to agree at leading order -- one-loop in Yang-Mills theory
-- and is close but not quite in agreement at order two Yang-Mills
loops. Furthermore, in addition to the leading 1/2 power in the
t'Hooft coupling, which is generic in string field theory, and which
we have previously argued cancels, we find that the 3/2 and 5/2
powers are also miraculously absent.}


\preprint{}

\begin{document}

\newcommand{\Tr}{\ensuremath{\mathop{\mathrm{Tr}}}}
\newcommand{\sign}{\ensuremath{\mathop{\mathrm{sign}}}}
\newcommand{\del}{\partial}
\def\be{\begin{equation}}
\def\ee{\end{equation}}
\def\bea{\begin{eqnarray}}
\def\eea{\end{eqnarray}}
\def\la{\langle}
\def\ra{\rangle}
\def\dag{\dagger}
\def\wt{\widetilde}
\def\wh{\widehat}

\def\da{\dot{\alpha}}
\def\db{\dot{\beta}}
\def\dg{\dot{\gamma}}
\def\dd{\dot{\delta}}
\def\dr{\dot{\rho}}
\def\ds{\dot{\sigma}}
\def\de{\dot{\epsilon}}

\def\lr{\leftrightarrow}
\def\be{\begin{equation}}
\def\ee{\end{equation}}
\def\bea{\begin{eqnarray}}
\def\eea{\end{eqnarray}}
\def\la{\langle}
\def\ra{\rangle}
\def\dag{\dagger}
\def\wt{\widetilde}
\def\wh{\widehat}

\def\da{\dot{\alpha}}
\def\db{\dot{\beta}}
\def\dg{\dot{\gamma}}
\def\dd{\dot{\delta}}
\def\dl{\dot{\lambda}}
\def\dr{\dot{\rho}}
\def\ds{\dot{\sigma}}
\def\de{\dot{\epsilon}}

\def\G{\Gamma}
\def\D{\Delta}
\def\L{\Lambda}
\def\S{\Sigma}
\def\a{\alpha}
\def\b{\beta}
\def\g{\gamma}
\def\k{\kappa}
\def\d{\delta}
\def\e{\varepsilon}
\def\m{\mu}
\def\tr{\mbox{tr}}
\def\n{\nu}
\def\s{\sigma}
\def\r{\rho}
\def\l{\lambda}
\def\t{\tau}
\def\o{\omega}
\def\O{\Omega}
\def\v{\varrho}
\def\vt{\vartheta}
\def\mc{\mathcal}
\def\N{\nabla}
\def\p{\partial}
\def\K{\widetilde{K}}
\def\tb{\tilde{b}}
\def\lr{\leftrightarrow}

%\numberwithin{equation}{section}

%\numberwithin{equation}{section}

%\end{center}
%\end{titlepage}
\section{Introduction and Conclusions}

The AdS/CFT correspondence~\cite{Juan,Gubser:1998bc,Witten:1998qj}
has provided one explicit example of the long conjectured duality
between gauge fields and strings. One of the most important testing
grounds for this correspondence is string theory in the pp-wave
geometry and its mapping to the BMN limit of Yang-Mills theory.

The pp-wave geometry is produced by taking the Penrose limit of
AdS$_5\times$S$^5$~\cite{Blau:2001ne,Blau:2002dy}. On that geometry,
non-interacting IIB string theory is explicitly solvable and the
complete spectrum of free strings can be found~\cite{Metsaev}. The
corresponding BMN limit of $\mathcal{N}=4$ super-Yang-Mills theory
can be taken by identifying the appropriate
operators~\cite{Berenstein:2002jq} and taking a large quantum number
limit. The planar limit of Yang-Mills theory corresponds to
non-interacting strings and the planar spectrum of the Yang-Mills
dilatation operator, which is dual to the string Hamiltonian, can be
computed
perturbatively~\cite{Berenstein:2002jq}-\cite{Santambrogio:2002sb}.
As far as these computations have been done, the result shows
beautiful agreement between planar Yang-Mills and non-interacting
strings. This agreement has been extended to scenarios beyond the
BMN limit~\cite{Gubser:2002tv,Callan:2003xr} and to the
non-perturbative sector~\cite{Green:2005pg,Green:2005rh} and  has
led to many promising insights.

One of those insights has been the recognition that the problem of
computing dimensions of composite operators in $\mathcal{N}=4$
super-Yang-Mills theory can be mapped onto integrable spin
chains~\cite{Minahan:2002ve}-\cite{Beisert:2003xu}. The string
theory sigma model on $AdS_5\times S^5$ also has an integrable
structure~\cite{Bena:2003wd} and much progress has been made to the
point that a complete matching of the precise details of planar
Yang-Mills and and non-interacting strings on the full $AdS_5\times
S^5$ background is a possibility that is sometimes
contemplated~\cite{Beisert:2005bm}-\cite{Rej:2005qt}.

However, in spite of this optimistic outlook, beyond the planar
limit of Yang-Mills theory and non-interacting string theory, there
has been very little success in checking the AdS/CFT correspondence,
even in the BMN limit.  For example, the Yang-Mills prediction for
the string-loop corrections to energies of 2-impurity BMN states
were computed early
on~\cite{Kristjansen:2002bb,Constable:2002hw,Beisert:2002bb,Constable:2002vq}.
The gauge theory prediction for the BMN energy of a 2-impurity state
is~\footnote{We remind the reader that the string light-cone momenta
are related to Yang-Mills conformal dimension $\Delta$ and R-charge
$J$ as
\begin{equation} p^-=\mu\left( \Delta-J\right) ~~,~~ p^+=\frac{
\Delta+J }{2 \mu \sqrt{ g_{YM}^2N } \alpha' }
\end{equation} where in the BMN limit $N,\Delta,J\to\infty$
so that $(p^+,p^-)$ remain finite. Two convenient couplings are
\begin{equation}
\frac{1}{\left(\mu\alpha'p^+\right)^2}=\frac{g_{YM}^2N}{J^2}
\equiv\lambda' ~~,~~ 4\pi g_s\left(\mu\alpha'p^+\right)^2 =
\frac{J^2}{N} \equiv g_2~~,~~N,J\to\infty
\end{equation}
$\lambda'$ is proportional to the string tension.  $g_2$ is the
string coupling which weights the genus of the string world-sheet. }

\begin{equation}\label{gauge}
\Delta-J =
2\left(1+{\small\frac{1}{2}}\lambda'n^2-{\small\frac{1}{8}}{\lambda'}^2n^4+\ldots
\right)+\frac{g_2^2}{4\pi^2} \left(\frac{1}{12}+\frac{35}{32\pi^2
n^2}\right)\left( \lambda'-\frac{1}{2}{\lambda'}^2 n^2\right)+\ldots
\end{equation}
Attempts to produce a result which matches this one using string
theory have spawned a large
literature~\cite{Spradlin:2002ar}-\cite{Dobashi:2006fu} the best
available published computation using light-cone string field theory
is due to Gutjahr and Pankiewicz~\cite{Gutjahr:2004dv} (their
Eq.~(4.17))
\begin{eqnarray}\label{string}
\frac{p^-}{\mu}=2\sqrt{1+\lambda'n^2}+
~~~~~~~~~~~~~~~~~~~~~~~~~~~~~~~~~~~~~~~~~~~~~~~~~~~~~~~~~~~~~~~~~~~~~~~~~~~~
\nonumber
\\ +\frac{g_2^2\lambda'}{4\pi^2}\left( \left(\frac{1}{24}+\frac{65}{64\pi^2
n^2}\right)  -\frac{3}{16\pi^2}{\lambda'}^{\small\frac{1}{2}}
-\frac{n^2}{2}\left(\frac{1}{24}+\frac{89}{64\pi^2n^2}\right){\lambda'}
+\frac{9n^2}{32\pi^2}{\lambda'}^{\small\frac{3}{2}}+\ldots\right)
\end{eqnarray}
This computation, as did those which preceded it, uses an
unjustified truncation of the string vertex to the 2-impurity
channel.  It clearly does not match the gauge theory result
(\ref{gauge}). The three-string vertex actually has an arbitrary
pre-factor, the choice of which gives an arbitrary re-scaling of the
entire expression in Eq.~(\ref{string}).  The pre-factor can thus be
chosen so that either the first or the second term in the leading
order $\lambda'$ contribution agrees with gauge theory, but not
both. There are other differences in the terms beyond the leading
order.

On the other hand, in spite of its shortcomings, the formula in
(\ref{string}) has some remarkable features.  The natural expansion
parameter on the string side is ${\lambda'}^{\small\frac{1}{2}}$. In
(\ref{string}), the naive leading term that one would expect from
power-counting, $\sim {\lambda'}^{\small\frac{1}{2}}$, is absent. It
was argued that this is generally so in Ref.~\cite{Grignani:2005yv}.
The leading non-zero term, of order $\lambda'$, has contributions of
the same functional form in $n$ as the gauge theory result, it is
only the coefficients that are wrong. The bigger problem begins with
the order ${\lambda'}^{\small\frac{3}{2}}$ term which is clearly
absent in the gauge theory, where the expansion parameter is in
integer powers of $\lambda'$. One might argue that such a fractional
power is generated non-perturbatively, by re-summing logarithmic
divergent diagrams for example, and it could appear in principle.
This does happen elsewhere, for example in the expansion of the free
energy of Yang-Mills theory at finite temperature. However, the
gauge theory result seems to be free of infrared problems, this has
been checked explicitly to at least order ${\lambda'}^2$, so it is
difficult to see how a term of order
${\lambda'}^{\small\frac{3}{2}}$ could occur.

In the present paper, we will repeat the light-cone string field
theory computation that led to (\ref{string}), using the same
truncation to the 2-impurity channel, and a modified form of the
pp-wave background string vertex which was suggested by Dobashi and
Yoneya in  Ref.~\cite{Dobashi:2004nm}. Other details of the
computation are identical to those in Ref.~\cite{Gutjahr:2004dv}
which led to (\ref{string}).\footnote{There are a few minor
corrections which affect the fractional powers in (\ref{string}),
but they remain non-zero in the corrected (\ref{string}).} Our
result will be
\begin{eqnarray}\label{ours}
\frac{p^-}{\mu}&=&2\sqrt{1+\lambda'n^2}+ \nonumber
\\ &+&\frac{g_2^2\lambda'}{4\pi^2}|f|^2 \left(\frac{1}{12}+\frac{35}{32\pi^2
n^2}\right)\left( \frac{3}{4}-\frac{n^2}{2}\lambda'+{\cal
O}({\lambda'}^2)\right)
\end{eqnarray}
where $f$ is the unknown pre-factor of the vertex.  Note that now,
remarkably, if we set the pre-factor $|f|^2=\frac{4}{3}$, the order
$\lambda'$ term agrees with gauge theory.  The order ${\lambda'}^2$
term, however, does not.  Further to this, the fractional powers of
$\lambda'$ are absent, at least up to order $7/2$.

 The essential new aspect of this computation is the use of the
Dobashi-Yoneya vertex. Unlike the case of Minkowski space, on the
pp-wave background there are competing proposals for the three
string vertex. The original one
\cite{Spradlin:2002ar,Spradlin:2002rv,Pankiewicz:2002gs,Pankiewicz:2002tg,Pankiewicz:2003kj}
(which we will call the SVPS vertex) was fixed using the
supersymmetry algebra up to a pre-factor function of the light-cone
momentum (which is $f$ in Eq.~(\ref{ours})). %One of the criterea
%used for finding the vertex was that it has a smooth limit as the
%pp-wave parameter $\mu$ goes to zero and the space-time metric
%approaches Minkowski space. If that requirement is relaxed, there is
%another vertex which is also a solution of the supersymmetry
%algebra.  It was proposed in Ref.~\cite{DiVecchia:2003yp} and we
Another vertex was proposed in Ref.~\cite{DiVecchia:2003yp} and we
will call it the DVPPRT vertex. The DVPPRT vertex solves the
supersymmetry algebra in the simplest possible way, by acting upon
the oscillator representation of the Dirac delta function which
enforces world-sheet locality by the {\it quadratic} Hamiltonian and
supercharge. This vertex is trivial in Minkowski space, but is
non-trivial in the pp-wave background.

Then, in Ref.~\cite{Dobashi:2004nm}, Dobashi and Yoneya proposed
another form for the cubic Hamiltonian and supercharge based on
consistency with the AdS/CFT holographic relations for three-point
functions~ \cite{Gubser:1998bc,Witten:1998qj}  and their comparison
with supergravity. This ``holographic'' vertex, which we shall call
the DY vertex, is an equal-weighted average of the original SVPS
vertex and the  DVPPRT vertex:
DY=${\small\frac{1}{2}}$SVPS+${\small\frac{1}{2}}$DVPPRT. It, and
the four-string contact term that is generated using the
supersymmetry algebra, are the vertices that are used in deriving
Eq.~(\ref{ours}). A correction to the DY vertex based on the
supersymmetry algebra was suggested in Ref.~\cite{Lee:2004cq}.
Because of the truncation to the 2-impurity channel, this
modification does not influence the computations in the present
paper.

 Though the result (\ref{ours}) is a big improvement on the previous one,
 it is still not in complete agreement with the gauge theory computation.
 It disagrees at order ${\lambda'}^2$. There might (or might not) be a
 simple reason for this disagreement.  We have not performed
 computations beyond the 2-impurity channel.  It could be that
 higher impurity channels contribute only to orders
 ${\lambda'}^2$ or higher, but do not influence the order $\lambda'$
 contribution.  This would require a miraculous cancelation of a number of orders
 in the small ${\lambda'}^{\small\frac{1}{2}}$ expansion.  After all, from power counting and
 the generic structure of the amplitude, one would expect that higher impurities
 begin to contribute at order ${\lambda'}^{\small\frac{1}{2}}$. In previous work,
 we have shown that this leading order cancels~\cite{Grignani:2005yv}. There, it was
 associated with cancelation of divergences, which were also generic, and supersymmetry
 played an important role. Examining whether this cancelation could also occur at orders
 $\lambda'$ and ${\lambda'}^{\small\frac{3}{2}}$ is a challenge
 that has not been addressed yet.  A careful check of this possibility
 would be very interesting.

 There is another possibility for discrepancy.  In all computations
 to date, the contact term with the supercharge $g_2^2Q_4$ has been
 assumed to not contribute.  Indeed, the supersymmetry algebra shows
 that the contact term is (schematically) $g_2^2H_4=g_2^2\{Q_3,Q_3\}
 + g_2^2\{Q_2,Q_4\} + g_2^2\{Q_4,Q_2\}$ and only the first term on the right-hand side
 has been used in all computations. Generally, these contact terms are needed to cancel
 divergences arising from iterations of lower order
 vertices~\cite{Greensite:1987hm,Green:1987qu}. In principle, $Q_4$ could be determined by
 finding multi-string matrix elements of the supersymmetry algebra.
 To our knowledge, this has not been attempted on the pp-wave
 background.
 We only observe that $Q_4$ is not needed to cancel divergences in any
 of the quantities that we compute. (This was also found on the Minkowski background
 in Ref.~\cite{Green:1987qu}.)  However, we cannot rule out
 its having a non-zero finite contribution that would affect our results.

 One further observation that we can make is that, we could consider
 any linear combination of the SVPS and DVPPRT vertices:
 $\alpha$SVPS+$\beta$DVPPRT.
 In this case, it would seem that, by using the supersymmetry algebra,
 one could consistently construct higher order
 contact terms in the Hamiltonian and supercharges, so that this is
 also a viable possibility for the vertex.  In particular, we will see that divergences
 in the energy shifts of 2-impurity states cancel for any values of $\alpha$ and $\beta$.
 However, if we use
 this vertex to compute the energy shift, we find a result that agrees with
 gauge theory (\ref{gauge}) to the leading order $\lambda'$ only for the
 particular combination in the DY vertex, that is only when
 $\alpha=\beta={\small\frac{1}{2}}$.

There is another intriguing and unexplained feature of these
results, which was observed in Ref.~\cite{Roiban:2002xr}. Consider
the expansion of the string field theory Hamiltonian into free
(quadratic) and interacting -- cubic, quartic, etc.~terms,
$H=H_2+g_2H_3+g_2^2H_4+\ldots$ and the expression for second order
quantum mechanical perturbation theory which is used to compute
(\ref{string}),
\begin{equation}\label{hhhhh} \delta E^{(2)} ~=~ g_2^2<\psi_0|
H_3\frac{1}{E_0-H_2}H_3|\psi_0>+g_2^2<\psi_0|H_4|\psi_0>
\end{equation}
If, in the computation which arrives at (\ref{string}), we change
the terms on the right-hand-side of (\ref{hhhhh}) by a relative
factor of 2, either multiplying the first term by $\frac{1}{2}$ or
the second term by $2$, then the order $\lambda'$ term would be
different from that quoted in (\ref{string}) and in that case the
pre-factor could be chosen so that the order $\lambda'$ term agrees
with gauge theory.   Here, we observe that this interesting fact
persists in (\ref{string}) to higher orders. In that case, with
factor of 2 and the same choice of prefactor the order $\l'^2$ term
also agrees with gauge theory, and the $\l'^{3/2}$ term vanishes. In
addition, this intriguing fact persists in the
computation of (\ref{ours}), % to a
%higher order. In that case, if one inserts the factor of 2, the
%pre-factor can be chosen so that the leading, order $\lambda'$ term
%still matches gauge theory, the next, order ${\lambda'}^2$ term also
%matches gauge theory and the fractional powers of $\lambda'$ still
%vanish to order 5/2.
if one inserts a relative factor of 2 in (\ref{hhhhh}), (\ref{ours})
is modified so that it agrees with gauge theory up to and including
order ${\lambda'}^2$ and the coefficients of the fractional powers
with exponents $3/2$ or $5/2$ still vanish. At this point, we have
no explanation for this fact. Inserting the factor of 2 is
definitely not mathematically correct here. Aside from the violence
it would do to quantum mechanical perturbation theory, it would
upset the divergence cancelation that was found in
Ref.~\cite{Grignani:2005yv}, for example. The reason, if any, for
this numerological coincidence remains a mystery.

In the remainder of this Paper, we will outline the computation that
leads to Eq.~(\ref{ours}).  The notation and techniques are
identical to those used in Ref.~\cite{Gutjahr:2004dv} and
Ref.~\cite{Grignani:2005yv} and we defer to them for the details.

\section{Divergence Cancelation}

The light-cone energy of the two-oscillator free string state on the
pp-wave background is $p^-=2\mu\sqrt{1+\lambda'n^2}$. This matches
the energy of the two impurity BMN operator in planar Yang-Mills
theory. The energy-shift of these states due to string loop
corrections is calculated in second order quantum mechanical
perturbation theory using the formula in Eq.~(\ref{hhhhh}). We will
call the first term in (\ref{hhhhh}) the ``$H_3$ term'' and the
second the ``contact term''.

In our previous paper \cite{Grignani:2005yv} we showed that, in the
computation of the energy-shifts of some two-impurity states using
SVPS vertex, the $H_3$ and contact terms individually contain
logarithmically divergent sums over intermediate state mode numbers.
These divergences were shown to always cancel, leaving a finite
result which leads as $g_2^2\l'$. This behavior was shown to be
generic, and to exist at arbitrary order in intermediate state
impurities.  This was important because, of course it is necessary
to obtain finite amplitudes.  In addition, it is also the mechanism
whereby the leading order ${\lambda'}^{\small\frac{1}{2}}$
contributions cancel.

We shall now show that this mechanism is at play for the DVPPRT
vertices, and that any linear combination of the SVPS and DVPPRT
vertices will similarly be divergence free. The special choice of an
equal weighted average - the DY vertex - is thus well behaved.

The simplest method to understand the divergence cancelation is to
consider the energy shift of the two-impurity trace state
\be\left|[{\bf 1}, {\bf 1}]\ra\right. =
\frac{1}{{2}}\alpha^{i\dagger}_n \alpha^{i \dagger}_{-n} |\alpha
\rangle\label{singlet}\ee restricted to the impurity conserving
channel. For details of this computation we refer the reader to
\cite{Grignani:2005yv} and for details of definitions and notation
to Ref.~\cite{Gutjahr:2004dv} and other literature quoted there. The
DVPPRT vertex is given by the following expressions
\cite{DiVecchia:2003yp},

\bea |H^D_3\ra =&&-
g_2\,f(\m\a_3\,,\,\frac{\a_1}{\a_3})\frac{\alpha'}{16\,\a_3^3}
\Bigl[K^2+\K^2-4Y^{\a_1\a_2}{\widetilde
Y}_{\a_1\a_2}-4Z^{\da_1\da_2}{\widetilde
Z}_{\da_1\da_2}\Bigr]|V\ra\,,\cr |Q^D_{3\,\b_1\db_2}\ra =&&
 g_2\,\eta\,f(\m\a_3\,,\,
\frac{\a_1}{\a_3})\frac{1}{4\, \a_3^3}\,\sqrt{-\frac{\a'\k}{2}}
\Bigl(Z_{\dg_1\db_2}K_{\b_1}^{\dg_1}- i
Y_{\b_1\g_2}K_{\db2}^{\g_2}\Bigr)|V\ra\,,\cr |Q^D_{3\,\db_1\b_2}\ra
=&& g_2\,\bar\eta\,f(\m\a_3\,,\, \frac{\a_1}{\a_3})\frac{1}{4\,
\a_3^3}\,\sqrt{-\frac{\a'\k}{2}}
\Bigl(Y_{\g_1\b_2}K_{\db_1}^{\g_1}-iZ_{\db_1\dg_2}K_{\b_2}^{\dg_2}\Bigr)|V\ra\,.
\label{divecchia} \eea

\noindent Unlike the SVPS case, the $H_3$ divergence does not stem
from the two-bosonic-impurity intermediate state. This can be traced
to the substitution of $K^2 + \wt K^2$ for $K\,\wt K$ in the $H_3$
prefactor. There is, however, another divergence that was not
present in the SVPS case. It is due to the contribution coming from
matrix elements with two fermionic impurities in the intermediate
state. In particular, the relevant matrix elements are given by

\bea\label{ferm} && \langle \alpha_3 | \alpha^i_{n} \alpha^i_{-n} \,
\langle \alpha_2 | \langle \alpha_1 | \beta_{p(1)}^{\a_1\a_2}
\beta_{-p(1)\b_1\b_2} |H^D_3 \rangle =\cr && 4\,g_2 r\,(1-r)\left(
\frac{\omega_{n}^{(3)}}{\alpha_3} +
\frac{\omega_{p}^{(1)}}{\alpha_1} \right)\left({\widetilde
Q}_{-p\,p}^{1\,1}\right. \left.- {\widetilde
Q}_{p\,-p}^{1\,1}\right){\widetilde
N}_{-n\,n}^{3\,3}\delta_{\b_1}^{\a_1} \delta_{\b_2}^{\a_2} \eea

\noindent and similarly for the intermediate state with dotted
indices. The divergent contribution to the energy shift coming from
these matrix elements is found by taking the large $p$ limits of the
summands in (\ref{hhhhh}). One finds

\begin{equation}\label{diveccH3div}
\delta E^{\mbox{div}}_{H^D_3} \sim -\frac{1}{2}\int_0^1 dr\,
\,\frac{g_2^2\,r(1-r)}{r\,|\alpha_3|\,\pi^2} \left( {\widetilde
N}^{3\,3}_{n\,-n} \right)^2 \, \sum_p \frac{1}{|p|}
\end{equation}

The contribution from the contact term stems from the following
matrix element,

\bea \label{contactmed} &&\left( g_2 \frac{\eta}{4}
\sqrt{\frac{r\,(1-r)\, \alpha'}{-2\,\alpha_3^3}} \right)^{-1}
\langle \alpha_3 | \alpha^i_{n} \alpha^i_{-n} \, \langle \alpha_2 |
\langle \alpha_1 | \alpha^{K\,(1)}_{p} \beta^{(1)\,
\Sigma_1\,\Sigma_2}_{-p} |Q^D_{3\, \beta_1 {\dot \beta}_2} \rangle
=\cr && 2\biggl( G_{|p|}^{(1)} \, K_{-n}^{(3)} {\widetilde
N}^{3\,1}_{n\,p} + G_{|p|}^{(1)} \, K_{n}^{(3)} {\widetilde
N}^{3\,1}_{-n\,p}
 \biggr)( \sigma^k)^{\dot{\sigma_1}}_{ \beta_1 } \delta^{{\dot \sigma_2}}_{{\dot \beta_2}}
+8\, G_{|p|}^{(1)} \, K_{p}^{(1)} {\widetilde N}^{3\,3}_{n\,-n} (
\sigma^K)^{\Sigma}_{ \beta } \delta^{\Sigma}_{\beta}. \eea

\noindent The divergent contribution to the energy shift is found to
be,

\begin{equation}
\delta E^{\mbox{div}}_{H^D_4} \sim +\int_0^1 dr\,
\,\frac{g_2^2\,r(1-r)}{r\,|\alpha_3|\,\pi^2} \left( {\widetilde
N}^{3\,3}_{n\,-n} \right)^2 \, \sum_{p>0} \frac{1}{p}
\end{equation}

\noindent Noting that in the $H^D_3$ contribution the divergence is
found for both positive and negative $p$, while in the $H^D_4$
contribution the divergence occurs only for negative $p$, and hence
a relative factor of 2 is induced in the $H^D_3$ term, one sees that
the logarithmically divergent sums cancel identically between the
$H^D_3$ and contact terms, leaving a convergent sum. This result can
be generalized to arbitrary impurity channels, as was done for the
SVPS case in \cite{Grignani:2005yv}.

We now show that an arbitrary linear combination of the SVPS and
DVPPRT vertices,

\bea H^N_3 = \alpha \,H^S_3 + \beta\, H^D_3 \\
Q^N_3 = \alpha\, Q^S_3 + \beta\, Q^D_3 \eea

\noindent similarly yields a finite energy shift. The divergence
stemming from the $H_3$ term is simply $\a^2$ times the SVPS $H_3$
divergence plus $\b^2$ times (\ref{diveccH3div}). The reason is
simple - the SVPS divergence stems from an entirely bosonic
intermediate state, while (\ref{diveccH3div}) results from an
entirely fermionic one. This precludes any divergences arising from
cross terms. Referring the reader to equation (2.7) of
\cite{Grignani:2005yv}, we note that the SVPS divergence is exactly
equal to (\ref{diveccH3div}), therefore we have,

\be \delta E^{\mbox{div}}_{H^N_3} \sim -(\a^2+\b^2)
\frac{1}{2}\int_0^1 dr\,
\,\frac{g_2^2\,r(1-r)}{r\,|\alpha_3|\,\pi^2} \left( {\widetilde
N}^{3\,3}_{n\,-n} \right)^2 \, \sum_p \frac{1}{|p|} \ee

\noindent The pieces of the SVPS $Q_3$ relevant to a two-impurity
channel calculation are exactly $Q_3^D$ with $K \lr \wt K$, see
again \cite{Grignani:2005yv}.

\bea \label{contactmey} &&\left( g_2 \frac{\eta}{4}
\sqrt{\frac{r\,(1-r)\, \alpha'}{-2\,\alpha_3^3}} \right)^{-1}
\langle \alpha_3 | \alpha^i_{n} \alpha^i_{-n} \, \langle \alpha_2 |
\langle \alpha_1 | \alpha^{K\,(1)}_{p} \beta^{(1)\,
\Sigma_1\,\Sigma_2}_{-p} |Q^Y_{3\, \beta_1 {\dot \beta}_2} \rangle
=\cr &&~~~ 2 G_{|p|}^{(1)} \,\biggl( \biggl[ \a
\,(K_{-n}^{(3)}{\widetilde N}^{3\,1}_{-n\,p}+K_{n}^{(3)}{\widetilde
N}^{3\,1}_{n\,p})+\b\,( K_{-n}^{(3)}{\widetilde N}^{3\,1}_{n\,p} +
K_{n}^{(3)}{\widetilde N}^{3\,1}_{-n\,p}) \biggr]
 ( \sigma^k)^{\dot{\sigma_1}}_{ \beta_1 } \delta^{{\dot \sigma_2}}_{{\dot
 \beta_2}}\cr
&&~~~~~+4\, (\b\,K_{p}^{(1)}+\a\,K_{-p}^{(1)}) {\widetilde
N}^{3\,3}_{n\,-n} ( \sigma^K)^{\Sigma}_{ \beta }
\delta^{\Sigma}_{\beta}\biggr) \eea

\noindent The last term in (\ref{contactmey}) gives rise to a
log-divergent sum, the large-$p$ behaviour of which is:

\begin{equation}
\delta E^{\mbox{div}}_{H^N_4} \sim +(\a^2+\b^2)\int_0^1 dr\,
\,\frac{g_2^2\,r(1-r)}{r\,|\alpha_3|\,\pi^2} \left( {\widetilde
N}^{3\,3}_{n\,-n} \right)^2 \, \sum_{p>0} \frac{1}{p}
\end{equation}

\noindent Thus the energy shift is finite for arbitrary $\a$ and
$\b$. The DY vertex uses $\a=\b=1/2$, and this combination
exclusively gives rise to the agreement with gauge theory discussed
in the introduction. The generalization of these arguments to the
impurity non-conserving channels is a straightforward application of
the treatment given in \cite{Grignani:2005yv}.

\section{Results}

The calculations undertaken in this Paper are practically identical
to those in \cite{Gutjahr:2004dv}, using the DVPPRT and DY vertices
in place of the SVPS vertices used there. One small difference in
the case of the SVPS vertex is that the half-integer powers of $\l'$
calculated in Ref.~\cite{Gutjahr:2004dv} and quoted vertabim in our
Eq.~(\ref{string}) are incomplete and suffer from a sign error, and
are correctly given below. We refer the reader to this reference for
details, and simply give results below.

The external state for which we are calculating the energy shift is

$$\left. |[{\bf 9}, {\bf
1}]\ra^{(ij)}\right. =
\frac{1}{\sqrt{2}}\left(\a^{\dag\,i}_n\a^{\dag\,j}_{-n}+\a^{\dag\,j}_n\a^{\dag\,i}_{-n}
-\frac{1}{2}\d^{ij}\a^{\dag\,k}_n\a^{\dag\,k}_{-n}\right)|3\ra.
$$

\noindent For this particular state, individual $H_3$ and contact
terms are not divergent in the two impurity approximation. It should
be further noted that for this state, and for the impurity
conserving channel, we shall find that use of the DY vertex, rather
than the SVPS vertex, is equivalent to making the replacements of
the quantities $(K,\wt K)$ as $K \rightarrow (K+\wt K)/2$ and $\wt K
\rightarrow (K+\wt K)/2$ in the SVPS vertex. This is the simplest
way of reproducing our results.

The separate $H_3$ and contact term contributions to the energy
shift for each of the three vertices are given below. We find that
the DY energy shift agrees with gauge theory only at the leading
order, while also enjoying the vanishing of the $3/2$ and $5/2$
powers of $\l'$. The order-$\l'^2$ term is of the correct form, but
suffers from an overall factor of $4/3$. The SVPS and DVPPRT results
do not agree with gauge theory at the leading order. By multiplying
the contact terms by two (an unjustified operation), one can recover
the correct gauge theory result up to $\l'^2$ order with the SVPS
(including vanishing of its $\l'^{3/2}$ term) and DY vertices.
Further, this operation does not spoil the vanishing $3/2$ and $5/2$
powers of $\l'$ for the DY result.

\subsection{$H_3$ terms}

\bea \d E^{\mbox{SVPS}}_{H_3} &=& \frac{g_2^2}{32\pi^2}\left[
\frac{15}{2\pi^2 n^2}\l'
+3\left(\frac{1}{\pi^2}+\frac{1}{2\pi}\right)\l'^{3/2}
-\frac{27}{4\pi^2}\l'^2
-n^2\left(\frac{5}{\pi^2}+\frac{9}{4\pi}\right)\l'^{5/2}\right.\cr
&+&\left.\frac{111 n^2}{16\pi^2}\l'^3
+n^4\left(\frac{45}{16\pi}+\frac{33\,}{5\pi^2}\right)\l'^{7/2}
+{\cal O}(\l'^4) \right] \eea

\bea \d E^{\mbox{DVPPRT}}_{H_3} &=& \frac{g_2^2}{32\pi^2}\left[
-\left(\frac{2}{3}+\frac{5}{4\pi^2 n^2}\right)\l'
+3\left(\frac{1}{\pi^2}+\frac{1}{2\pi}\right)\l'^{3/2}
+n^2\left(1-\frac{9}{8\pi^2 n^2}\right)\l'^2 \right.\cr &-& \left. 5
n^2\left(\frac{2}{\pi^2}+\frac{3}{4\pi}\right)\l'^{5/2}
-5n^4\left(\frac{1}{4}-\frac{21}{32\pi^2 n^2}\right)\l'^3
+n^4\left(\frac{105}{16\pi}+\frac{94\,}{5\pi^2}\right)\l'^{7/2}
\right.\cr &+& \left.{\cal O}(\l'^4) \right]\eea

\bea \d E^{\mbox{DY}}_{H_3} &=& \frac{g_2^2}{4\pi^2}\left[
\frac{3}{4}\left(\frac{1}{12}+\frac{35}{32\pi^2 n^2}\right)\l'
-5n^2\left(\frac{1}{96}+\frac{35}{256\pi^2 n^2}\right)\l'^2
\right.\cr &+& \left. n^4\left(\frac{17}{384}+\frac{655}{1024\pi^2
n^2}\right)\l'^3
+n^4\left(\frac{3}{256\pi}+\frac{23\,}{640\pi^2}\right)\l'^{7/2}
+{\cal O}(\l'^4) \right] \eea

\subsection{Contact terms}

\bea \d E^{\mbox{SVPS}}_{H_4} &=& \frac{g_2^2}{32\pi^2}\left[ \left(
\frac{1}{3}+\frac{5}{8\pi^2
    n^2}\right) \l'
-\frac{3}{2}\left(\frac{1}{\pi^2}+\frac{1}{2\pi}\right)\l'^{3/2}
-n^2\left(\frac{1}{6}-\frac{19}{16\pi^2 n^2}\right)\l'^2 \right. \cr
&+& \left.n^2\left(\frac{11}{4\pi^2}+\frac{9}{8\pi}\right)\l'^{5/2}
+\frac{n^4}{8}\left(1 - \frac{105}{8\pi^2 n^2}\right)\l'^3
-n^4\left(\frac{45}{32\pi}+\frac{73\,}{20\pi^2}\right)\l'^{7/2}
\right. \cr &+& \left.{\cal O}(\l'^4) \right] \eea

\be \d E^{\mbox{DVPPRT}}_{H_4} = \d E^{\mbox{SVPS}}_{H_4} \ee

\bea \d E^{\mbox{DY}}_{H_4} &=& \frac{g_2^2}{4\pi^2}\left[
n^2\left(\frac{1}{96}+\frac{35}{256\pi^2 n^2}\right)\l'^2 -\frac{5
n^4}{128} \left(\frac{1}{3}+\frac{29}{8\pi^2 n^2}\right)\l'^3
\right.\cr &+&
\left.\frac{n^4}{256}\left(\frac{3}{2\pi}+\frac{5}{\pi^2}\right)\l'^{7/2}
+{\cal O}(\l'^4) \right] \eea

\subsection{Energy shifts}

The results for the complete energy shifts are as follows,

\bea \d E^{\mbox{SVPS}} &=&  \frac{g_2^2}{4\pi^2}\left[
\left(\frac{1}{24} + \frac{65}{64\pi^2 n^2}\right)\l'
+\frac{3}{16}\left(\frac{1}{\pi^2}+\frac{1}{2\pi}\right)\l'^{3/2}\right.\cr
&-& \left. n^2\left(\frac{1}{48}+\frac{89}{128\pi^2 n^2}\right)\l'^2
-\frac{9\,n^2}{32}\left(\frac{1}{\pi^2}+\frac{1}{2\pi}\right)\l'^{5/2}
\right. \cr &+& \left.n^4\left(\frac{1}{64}+\frac{339}{512\pi^2
n^2}\right)\l'^{3} + n^4 \left( \frac{59}{160\pi^2} +
\frac{45}{256\pi}\right)\l'^{7/2} +{\cal O}(\l'^4) \right] \eea

\bea \d E^{\mbox{DVPPRT}} &=&  \frac{g_2^2}{4\pi^2}\left[
-\left(\frac{1}{24} + \frac{5}{64\pi^2 n^2}\right)\l'
+\frac{3}{16}\left(\frac{1}{\pi^2}+\frac{1}{2\pi}\right)\l'^{3/2}\right.\cr
&+& \left.n^2\left(\frac{5}{48}+\frac{1}{128\pi^2 n^2}\right)\l'^2
-n^2\left(\frac{29}{32\pi^2}+\frac{21}{64\pi}\right)\l'^{5/2}\right.\cr
& +&\left. n^4\left(-\frac{9}{64}+\frac{105}{512\pi^2
n^2}\right)\l'^{3} + n^4 \left( \frac{303}{160\pi^2} +
\frac{165}{256\pi}\right)\l'^{7/2} +{\cal O}(\l'^4) \right] \eea

\bea\label{N4final} \d E^{\mbox{DY}} &=& \frac{g_2^2}{4\pi^2}
\frac{3}{4} \left[ \left(\frac{1}{12}+\frac{35}{32\pi^2 n^2}\right)
\left(\l'-\frac{4}{3}\frac{n^2}{2}\l'^2\right)
+\frac{n^4}{24}\left(1+\frac{255}{16\pi^2 n^2}\right)
\l'^3\right.\cr &+&\left.
\frac{n^4}{384}\left(\frac{9}{\pi}+\frac{142}{5\pi^2}\right)\l'^{7/2}
+{\cal O}(\l'^4) \right] \eea

\noindent Recall that the leading $3/4$ is irrelevant and can be
scaled away by fixing the overall $f$ factor which multiplies the
vertices (and which has not been written in the above formulae,
where it would appear in each as an overall factor of $|f|^2$). We
see that the gauge theory result (\ref{gauge}) is matched only by
the DY result, and only at leading order in $\l'$, with the $\l'^2$
term being of the correct form but with an overall factor of $4/3$.
We also see the miraculous absence of the $\l'^{3/2}$ and
$\l'^{5/2}$ terms which are clearly generic in the string field
theory. The result (\ref{N4final}) represents the best matching of
this quantity to gauge theory so far, and thus is an indication that
the DY vertex is an improvement over its predecessors.

Mysteriously, if the contact terms are scaled by a factor of 2, the
agreement with gauge theory is enhanced for both the SVPS and DY
results,

\bea \d E^{\mbox{SVPS}}_{2H_4} &=& \frac{g_2^2}{4\pi^2} \left[
\left(\frac{1}{12}+\frac{35}{32\pi^2 n^2}\right)
\left(\l'-\frac{n^2}{2}\l'^2\right) +
\frac{n^2}{16\pi^2}\l'^{5/2}\right.\cr &+&
\left.n^4\left(\frac{1}{32}+\frac{117}{256\pi^2 n^2}\right) \l'^3
-\frac{7 n^4}{80\pi^2}\l'^{7/2} +{\cal O}(\l'^4) \right] \eea

\bea \d E^{\mbox{DY}}_{2H_4} &=& \frac{g_2^2}{4\pi^2}
\frac{3}{4}\left[ \left(\frac{1}{12}+\frac{35}{32\pi^2 n^2}\right)
\left(\l'-\frac{n^2}{2}\l'^2\right)
+n^4\left(\frac{7}{288}+\frac{365}{768\pi^2 n^2}\right)
\l'^3\right.\cr &+& \left.
n^4\left(\frac{1}{10\pi^2}+\frac{1}{32\pi}\right)\l'^{7/2}+{\cal
O}(\l'^4) \right] \eea

\noindent however, the DY result is still superior in that the
$\l'^{5/2}$ power is absent.

\section*{Acknowledgements:} The work of B.~Ramadanovic, G.~Semenoff
and D.~Young is supported by the Natural Sciences and Engineering
Research Council of Canada and by University of British Columbia
Graduate Fellowships.  G.~Semenoff acknowledges the hospitality of
the University of Perugia where some of this work was done.  The
work of M.~Orselli is supported in part by the European Community's
Human Potential Programme under contract MRTN-CT-2004-005104
`Constituents, fundamental forces and symmetries of the universe'.
The work of G.Grignani is supported by the I.N.F.N.~ and M.U.I.R.~of
Italy. G.~Grignani and M.~Orselli acknowledge the hospitality of the
Pacific Institute for Theoretical Physics and the University of
British Columbia where parts of this work were done.

\end{document}